\documentstyle[twoside,fleqn,epsfig,espcrc2]{article}

\newcommand{\AmS}{{\protect\the\textfont2
  A\kern-.1667em\lower.5ex\hbox{M}\kern-.125emS}}

\title{Search for magnetic monopoles with nuclear track detectors}

\author{M. Giorgini\address{Dipartimento di 
Fisica, Universit\`a di Bologna
 and INFN, Sezione di Bologna \\ 
Viale C. Berti Pichat 6/2, 40127 Bologna, Italy \\
e-mail: miriam.giorgini@bo.infn.it} for the MACRO Collaboration\thanks{For 
        the complete list of the Collaboration see the paper
	by L. Patrizii in these Proceedings.}}
       
\begin{document}

\begin{abstract}
This paper describes an experimental search for GUT magnetic monopoles 
in the MACRO experiment using the nuclear track subdetector CR39. After 
discussing
the working principle, the charge resolution and
 the calibration     
of the detector, the experimental procedure for searching for
magnetic monopoles is described. Since no candidates were found, 
 the upper flux limits obtained by the MACRO CR39 used
as a ``stand alone detector" for 
magnetic monopoles of different magnetic charges are presented. 
\end{abstract}

\maketitle

\section{INTRODUCTION}
\vspace{-0.25cm}
Massive magnetic monopoles ($M >10^{16}$ GeV) are predicted by all
Grand Unified Theories of the electroweak and strong interactions 
(GUTs) \cite{Thooft}. They could carry a magnetic charge 
$g=ng_D$, where $n$ is an integer and $g_D$ is the elementary magnetic charge
 predicted by Dirac \cite{Dirac}. They should have been 
produced very early in the Big Bang Universe and should have
cooled to very low velocities since then. \par
MACRO (Monopole, Astrophysics and Cosmic Ray Observatory) \cite{Macro} 
is a large area 
underground experiment operating at Laboratori Nazionali del Gran Sasso,
 Italy, at an average depth of 3700 hg/cm$^2$. It has a
 modular structure and three types of subdetectors: streamer 
tubes, liquid scintillators and nuclear track detectors (CR39 and Lexan). \par
The track-etch
system is distribuited on three ortogonal planes
for a total area of 1263 m$^2$. \par
 In this paper the calibration, the charge resolution and the results
of the magnetic monopole (MM) search using the MACRO CR39 as a ``stand alone 
detector" are discussed. In particular, the attention 
is focused on the experimental
procedure of the search for magnetic monopoles: the etching conditions and
 the methods of searching for candidate events. In absence of candidates, 
 an updated estimate of the upper limits for an isotropic flux of 
monopoles with magnetic charge $g=g_D,~2g_D,~3g_D$ and for 
monopole-proton (M+p) composites established
by the MACRO nuclear track detector is also presented.

\vspace{-3mm}
\section{THE CR39 NUCLEAR TRACK DETECTOR}
\vspace{-0.2cm}
The MACRO CR39 is manufactured by the Intercast Europe Company of
Parma (Italy). The standard CR39, used mainly for sun glasses, was
improved in order to achieve a lower detection threshold, a higher 
sensitivity in a large range of energy losses, a high quality of the
post-etched surfaces after prolonged etching, stability of the sensitivity 
of the polymer over long periods of time and uniformity of 
sensitivity \cite{Produz}.

\vspace{-0.1cm}

\subsection{Working principle}
\begin{figure}[ht]
\vspace{-1cm}
\begin{center}
\mbox{
\epsfig{file=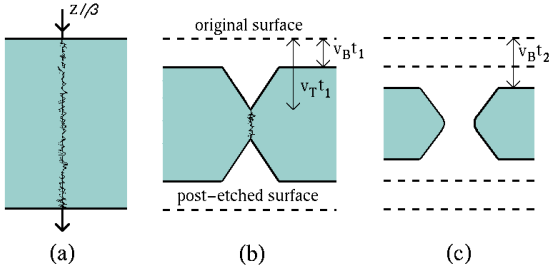,height=3.5cm}}
\end{center}
\vspace{-1cm}
\caption{\small 
(a) In a polymeric nuclear track detector, a heavily ionizing
particle breaks the polymeric bonds along its path, leaving
reactive molecules; (b) a chemical etching in a basic water solution
acts globally with velocity  $v_B$ and preferentially along the path
of the particle with velocity  $v_T$, forming, after the time $t_1$,
 observable cones on the post-etched surfaces and, after $t_2$, a hole (c).}
  \label{fig:funzionamento}
\vspace{-0.8cm}
\end{figure}

When a heavily ionizing particle crosses a nuclear track detector 
foil, it produces damages at the
level of molecular bonds, forming the so called ``latent track". During the
chemical etching of the detector in a basic water solution, as the 
etching velocity along the ``latent track",
 $v_T$, is larger than that for the bulk material, $v_B$,  etch-pit
cones are formed on both sides of the foil, see 
Fig. \ref{fig:funzionamento}. The  base area and the
height of each cone are functions of the Restricted Energy Loss (REL) of
the incident ion and thus of the charge $Z$ \cite{Fpw,Ncim}. 

\vspace{-0.1cm}
\subsection{Charge resolution of the CR39 detector}
\label{par:risoluzione}
 Stacks of CR39-target-CR39 were exposed to a Pb$^{82+}$ ion beam of 158
A GeV at the CERN-SPS. The exposures were performed at normal 
incidence. The 
beam passed through some foils of CR39 detector, interacted in the target 
material and then passed through CR39 foils which recorded the surviving
lead projectiles and their fragments.
After exposure, the CR39 detectors were etched for 268 h
in a 6N NaOH water solution at a temperature of 45 $^\circ$C. \par
\begin{figure}[htb]
\vspace{-1cm}
\begin{center}
\mbox{
\epsfig{file=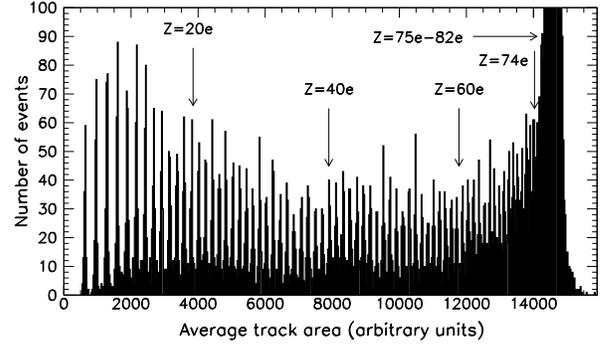,height=4.5cm}}
\end{center}
\vspace{-1cm}
\caption{\small Distribution of etched cone base areas averaged
on $12 \div 14$ faces of the CR39 sheets located after the fragmentation 
target. This
procedure allows to separate all single
peaks up to $Z=74e$; 
 for $Z \geq 75e$ the fragment tracks and the lead tracks cannot be
distinguished.}
  \label{fig:area}
\vspace{-0.7cm}
\end{figure}

 For each etch-pit cone, the base area 
was measured with an automatic
image analyzer system. In order to reconstruct the path of
the ions, we used a tracking procedure and then we performed for each ion
an average of the measured areas.
 A distribution of etched cone base areas
averaged on $12 \div 14$ faces of the CR39 sheets located
after the fragmentation
target is shown in Fig. \ref{fig:area}. This
procedure has a good charge resolution for $Z \leq 60e$ and an
acceptable one up to $Z \sim 74e$;
 for $Z \geq 75e$ the fragment tracks and the lead tracks cannot be
distinguished.
 The charge resolution at $Z\sim 10e$ is $\sim 0.3e$
for a single measurement and $\sim 0.09e$ for the average of 12
measurements \cite{Sezion}. The charge resolution is worsening with
 increasing charge of the ions.

\par
\begin{figure}[htb]
\vspace{-0.7cm}
\begin{center}
\mbox{
\epsfig{file=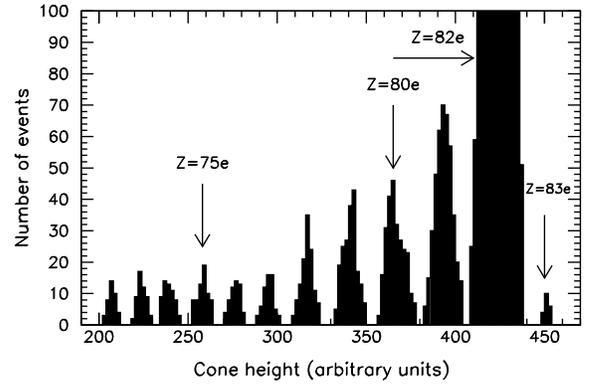,height=5cm}}
\end{center}
\vspace{-1cm}
\caption{\small Distribution of etched cone heights in a CR39 sheet 
located after the fragmentation target.}
  \label{fig:coni}
\vspace{-0.8cm}
\end{figure}

 For high $Z$ nuclei, the height of the etched cone is
more sensitive to $Z$ than its base area or diameter \cite{Nim}.
 In order to separate the surviving lead ions from the nuclear fragments
with charge $Z \geq 75e$,  we performed manual
 cone height measurements  with an optical microscope.
 The charge resolution obtained from this measurement technique is 
about $0.19e$ on a 
single face of the detector, see Fig. \ref{fig:coni}.

\subsection{Calibration of CR39 detectors}
This work was aimed at establishing the dependence of the response
of the CR39 nuclear track detector on the energy losses of charged particles.
 The calibration curve of a nuclear track detector is
the plot of the reduced etch rate, $p=v_T/v_B$, versus REL. For an 
ion passing through a nuclear track detector, REL 
is the fraction of the energy loss localized in a cilindrical region with 
diameter of 100 \AA ~around its path. It is the 
energy loss which leads to $\delta$ rays with energies lower than 
$E_{max}$, where $E_{max} \simeq 200$ eV for CR39. \par
\begin{figure}[htb]
\vspace{-1cm}
\begin{center}
\mbox{
\epsfig{file=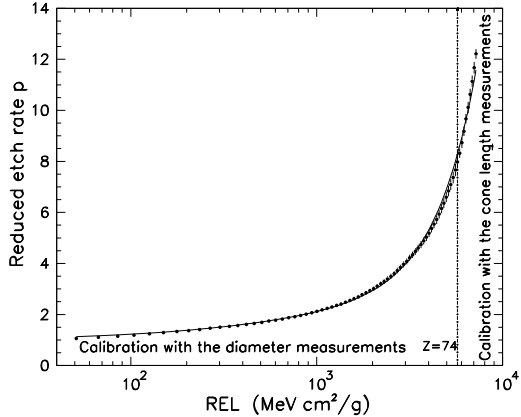,height=5.5cm}}
\end{center}
\vspace{-1cm}
\caption{\small  Reduced etch rate $p$ vs REL for the MACRO CR39
exposed to a relativistic lead beam; the points are the 
experimental data, the
solid line is the best fit to the data points.} 
  \label{fig:ext-cal}
\vspace{-0.7cm}
\end{figure}

After the exposure described in \S~\ref{par:risoluzione}, using 
measurements of the 
post-etched cone base areas and
heights in the charge regions $7e\leq Z \leq 74e$ and $75e < Z \leq 83e$,
 respectively, we obtained the calibration curve shown in
Fig. \ref{fig:ext-cal}. 

\par The comparison of this result  with  previous calibrations 
using relativistic and low velocity 
ions \cite{Ncim} shows that a unique curve  
describes, within errors, all the data. We can conclude that the 
response of the detector depends only from REL.   

\section{THE MACRO TRACK-ETCH SYSTEM}
\vspace{-0.2cm}
The MACRO track-etch detector consists of two types of 
nuclear track detectors, CR39 and Lexan. Most of the work concerns the 
development and processing of the CR39. Lexan has a much higher threshold
compared to that of CR39, making it sensitive to relativistic 
monopoles only. Lexan has not been used until now for monopole searches in 
MACRO.  \par
\begin{figure}[ht]
\vspace{-0.7cm}
\begin{center}
\mbox{
\epsfig{file=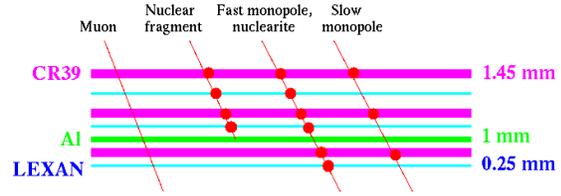,height=2.6cm}}
\end{center}
\vspace{-1cm}
\caption{\small Stack of nuclear track detectors used in MACRO. The black 
points indicate the breaking of polymeric bonds forming the ``latent track".}
  \label{fig:vagone}
\vspace{-0.7cm}
\end{figure}

The MACRO track-etch detector is organized 
in stacks (``wagons") consisting of 
three layers of CR39, each about 1.4 mm thick, three layers of Lexan, each 
0.25 mm thick, and an aluminium absorber 1 mm 
thick, Fig. \ref{fig:vagone}. Each stack has 
a surface of $24.5 \times 24.5$ cm$^2$ and is placed 
in an aluminized mylar bag filled with dry air. 

\par The main purpose of the track-etch detector is to offer independent
measurements of MMs. It may be used both as a ``stand alone 
detector" and ``triggered" by other MACRO subdetectors. 

\subsection{The response of CR39 to monopoles}
\begin{figure}[ht]
\vspace{-0.7cm}
\begin{center}
\mbox{
\epsfig{file=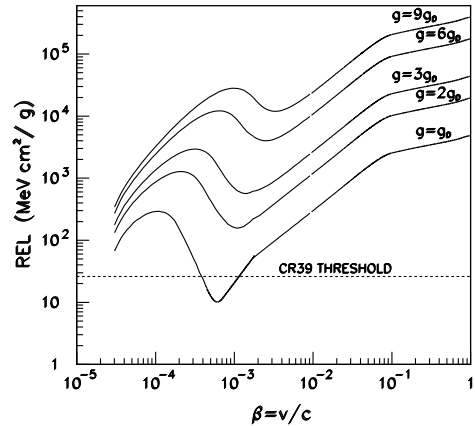,height=5.6cm}}
\end{center}
\vspace{-1cm}
\caption{\small REL vs $\beta$ for magnetic monopoles of different 
magnetic charge $g$.}
  \label{fig:rel}
\vspace{-0.7cm}
\end{figure}

The Restricted Energy Loss  as a function of the
monopole velocity in CR39 for 
$g=g_D,~2g_D,~3g_D,~6g_D,~9g_D$ bare monopoles is shown in Fig. \ref{fig:rel}
\cite{Rel}.
 The horizontal dashed line represents the detection threshold for the
MACRO CR39, corresponding to the REL value of $\sim 26$ MeV 
cm$^2$ g$^{-1}$ and to $Z/\beta \sim 5$. 

\par  The detection of MMs is limited to poles with
incidence angles smaller than a critical
value $\delta_c$ with respect to the normal; one has
$\delta_c = \arccos~(p^{-1})$, where $p = v_T/v_B$ \cite{Fpw}. \par
In order  to determine the 
acceptance of the MACRO CR39, we performed a Monte Carlo simulation,
assuming an isotropic flux of monopoles. This
requires that the masses of the monopoles should
be greater than $10^{17}$ GeV and their velocities should be larger than
$\beta = 3 \times 10^{-5}$ for $g = g_D$ monopoles and $ \beta = 5 \times
10^{-5}$ for $g = 2g_D,~3g_D$  monopoles and (M+p) aggregates.  The total
acceptance of the MACRO CR39 subdetector is 7100 m$^2$sr
for an isotropic flux of fast monopoles. 

\subsection{Experimental procedure and results}
The passage of a MM in CR39 is expected to cause
a structural damage. Chemical etching should result in collinear etch-pit
cones of equal size on both faces of each sheet. After extraction from the 
MACRO apparatus, three fiducial holes are drilled in each wagon by a 
precision drilling machine. The  
next steps are: (i) a ``strong"
etching in 8N NaOH water solution at 85 $^{\circ}$C for 50 hours is applied
to the first upper sheet of each wagon; (ii)  after etching, we perform
a first scan in transparency  
and a successive analysis with a binocular microscope with low magnification 
($16 \times$), looking for holes and/or corresponding double etch-pits;
 (iii) in case of a candidate in the upper sheet, a ``standard" etching 
in 6N NaOH water solution at 70 $^{\circ}$C for 30 hours is applied 
to the third sheet. A good candidate track must 
satisfy a three-fold coincidence
of the position, incident angles among the layers and should also give
the same value of REL. \par
\begin{figure}[ht]
\vspace{-0.7cm}
\begin{center}
\mbox{
\epsfig{file=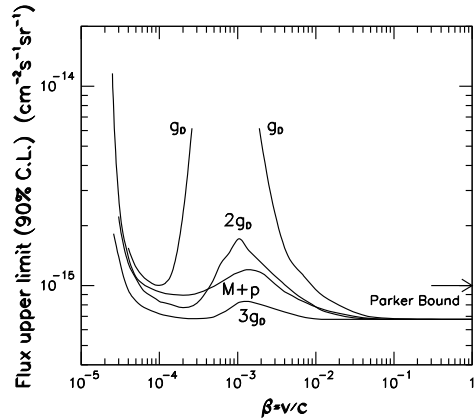,height=5.5cm}}
\end{center}
\vspace{-1cm}
\caption{\small 90\% C.L. upper limits for magnetic monopoles
obtained using MACRO CR39 as a ``stand alone detector".} 
  \label{fig:limite}
\vspace{-0.8cm}
\end{figure}

From 1991 until now we etched 227 m$^2$ of CR39, with an average 
exposure time of 7.6 years. Since no candidates were found,  the
90\% C.L. upper limits are given by $\Phi <
2.3/(S \Omega \cdot \Delta t)$ where $ S \Omega$  is the acceptance of the
extracted wagons and $\Delta t$ is their exposure time. Since the
limits corresponding to each individual extraction are independent,
 the resulting MACRO CR39 limit is
given by the sum of each individual $ (S\Omega \cdot \Delta t)$ 
contribution. In 
Fig. \ref{fig:limite} we present the flux upper limits for 
$g=g_D,~2g_D,~3g_D$
 bare MMs and for M+p composites 
obtained by our analysis. 

\subsection{Byproducts of the search for MMs}
The searches for magnetic monopoles based on the nuclear track detectors 
may also be applied to search for other rare particles. \par
 Nuclearites are hypothesized nuggets of strange quark matter and 
possible candidates for the dark matter. The CR39 detector may 
detect nuclearites
with $\beta > 10^{-5}$. In absence of candidates,  the 90\% C.L.
upper limit for an isotropic flux  
of nuclearites with $\beta > 10^{-5}$ is discussed in refs. \cite{Nuclea}. \par
Q-balls should be supersymmetric coherent states of squarks, sleptons 
and Higgs fields. The CR39 detector may detect charged Q-balls 
with $10^{-4} < \beta < 10^{-2}$.

\end{document}